\documentclass[aps,superscriptaddress,twocolumn,showpacs,prl,floatfix]{revtex4} %% groupedaddress,
\usepackage{graphicx,rotating,subfigure,amsmath,amsfonts,amssymb,delarray}

\newcommand{\e}{\text{e}}
\newcommand{\im}{\text{i}}
\def\l{\left}
\def\r{\right}
\def\12{\frac{1}{2}}

\begin{document}
% You should use BibTeX and apsrev.bst for references
\bibliographystyle{apsrev}

%\preprint{}

\title{Chain breaks and the susceptibility of
  Sr$_2$Cu$_{1-x}$Pd$_x$O$_{3+\delta}$ and
other doped quasi one-dimensional antiferromagnets}

%% Magnetic susceptibility of Sr$_2$Cu$_{1-x}$Pd$_x$O$_3$ and other doped
%%  one-dimensional antiferromagnets}
%% Curie-like behavior of the susceptibility in doped one-dimensional
%%  antiferromagnets}
%% Magnetic susceptibility for doped one-dimensional antiferromagnets
% Optional argument for running titles on pages
%\title[]{}


% other informatio
\author{J. Sirker}
\affiliation{Department of Physics and Astronomy, University of British
  Columbia, Vancouver, British Columbia, Canada V6T 1Z1}
\author{N. Laflorencie}
\affiliation{Department of Physics and Astronomy, University of British
  Columbia, Vancouver, British Columbia, Canada V6T 1Z1}
\author{S. Fujimoto}
\affiliation{Department of Physics, Kyoto University, Kyoto 606-8502, Japan}
\author{S. Eggert}
\affiliation{Department of Physics, University of Kaiserslautern, D-67663
  Kaiserlautern, Germany}
\author{I. Affleck}
\affiliation{Department of Physics and Astronomy, University of British
  Columbia, Vancouver, British Columbia, Canada V6T 1Z1}

\date{\today}

\begin{abstract}
  We study the magnetic susceptibility of one-dimensional $S=1/2$
  antiferromagnets containing non-magnetic impurities which cut the chain into
  finite segments. For the susceptibility of long anisotropic Heisenberg
  chain-segments with open boundaries we derive a parameter-free result at low
  temperatures using field theory methods and the Bethe Ansatz. The analytical
  result is verified by comparing with Quantum-Monte-Carlo calculations. We
  then show that the partitioning of the chain into finite segments can
  explain the Curie-like contribution observed in recent experiments on
  Sr$_2$Cu$_{1-x}$Pd$_x$O$_{3+\delta}$. Possible additional paramagnetic impurities
  seem to play only a minor role.
\end{abstract}
\pacs{75.10.Jm, 75.10.Pq, 02.30.Ik}
%%\pacs{05.10.Cc}
%%\pacs{05.70.-a}
%\keywords{}

\maketitle

%% Measurements of the magnetic susceptibility $\chi$ on antiferromagnets at low
%% temperatures $T$ often reveal a term $\chi\sim 1/T$. This Curie-like
%% contribution is usually associated with paramagnetic impurities (free spins).
%% For one-dimensional $S=1/2$ systems, however, a second mechanism exists which
%% can lead to such a contribution: Non-magnetic impurities or structural defects
%% can cut the chain into finite segments with essentially free boundaries. At
%% low temperatures $T/J\ll 1/L$ ($J$ being the exchange constant) a segment with
%% odd length $L$ will lock into its doublet ground state yielding also a
%% $1/T$-contribution \cite{EggertAffleck92}. As has been shown recently, a
%% $(T\ln T)^{-1}$-contribution even exists in the opposite limit $1/L \ll T/J$
%% due to the
%% open boundaries %% although suppressed by a power of $1/L$
%% \cite{FujimotoEggert,BortzSirker,SirkerBortzJSTAT}.

Measurements of the magnetic susceptiblity, $\chi$, on antiferromagnetics at
low temperatures, $T$, often reveal a (Curie) term $\chi \propto 1/T$.  This
is usually associated with paramagnetic impurities (free spins). For
one-dimensional S=1/2 systems, however, a second mechanism exists which leads
to a contribution with somewhat similar $T$-dependence.  Non-magnetic
impurities %% or structural defects 
can cut the chain into finite segments with essentially free boundaries. At
$T/J\ll 1/L$ ($J$ being the exchange constant) a segment with odd length $L$
will lock into its doublet ground state yielding also a $1/T$ contribution
\cite{BonnerFisher,EggertAffleck92,WesselHaas,SchmidtYushankhai}. However, the
behavior is considerably more complicated
at higher $T$. %%\cite{FujimotoEggert,BortzSirker,SirkerBortzJSTAT}
 As we show, fitting data to a Curie form for a limited range of $T$ can lead
to an underestimate of the impurity concentration by as much as a factor of
ten! In general, the different $T$-dependence %% for chain breaks 
could make it possible to distinguish this type of impurity from paramagnetic
ones providing useful structural information.

These results are relevant for materials like Sr$_2$CuO$_{3+\delta}$ which is
known to be an almost ideal realization of the spin-$1/2$ Heisenberg chain
with a nearest-neighbor coupling constant $J\sim 2200$ K and a low N\'eel
temperature $T_N\sim 5$ K$\sim 0.002\, J$ \cite{AmiCrawford,MotoyamaEisaki}.
%% suggesting that for temperatures $T\gg T_N$ the interchain couplings can be
%% ignored.
%% Furthermore, longer range couplings within the chain also seem to be very
%% small. 
Measurements of the magnetic susceptibility %% for this system 
have revealed a Curie contribution which could be dramatically reduced by
annealing. The Curie term therefore is believed to be mainly caused by excess
oxygen \cite{MotoyamaEisaki,AmiCrawford}.
%% This indicates that this contribution cannot be dominated by magnetic
%% impurities and that the segmentation
%% of the chain due to structural defects %% or non-magnetic impurities plays 
%% is more important. 
Recently, susceptibility measurements have also been reported on
Sr$_2$Cu$_{1-x}$Pd$_x$O$_{3+\delta}$ with Pd serving as a non-magnetic impurity
\cite{KojimaYamanobe}. 
%% The latter experiment has shown that the Curie-like
%% contribution at low $T$ increases monotonically with increasing Pd
%% concentration $x$.

In this letter we show that these susceptibility data can be explained by
taking the segmentation of the chain due to the non-magnetic impurities into
account properly {\it without} including any additional paramagnetic
impurities. To this end we derive a {\it parameter-free} analytical result for
the susceptibility of an anisotropic Heisenberg chain with finite length and
open boundary conditions (OBCs) at low temperatures using field-theory
methods.  This extends and generalizes previous results in the scaling limit
\cite{EggertAffleck92,WesselHaas} and for the boundary susceptibility in the
thermodynamic limit
\cite{FujimotoEggert,BortzSirker,SirkerBortzJSTAT,FurusakiHikihara}. We will
verify our analytical result using Quantum-Monte-Carlo (QMC) calculations.

We consider an anisotropic spin-$1/2$ Heisenberg chain consisting of segments
with length $L$ and OBCs
\begin{equation}
\label{Hseg}
H = J \sum_{i=1}^{L-1} \l[S^x_i S^x_{i+1} + S^y_i S^y_{i+1} +
\Delta S^z_i S^z_{i+1} \r] \; .
\end{equation}
Here, $\Delta$ parameterizes the exchange anisotropy. We assume that $T\gg T_N$
so that interchain couplings can be safely ignored. As argued at the end of
this letter, longer-range interactions which could bridge between the chain
segments can be neglected as well in the parameter regime we are interested
in. For a concentration of chain breaks $p$
%% the probability of finding a segment of length $L$ is then given by $p(1-p)^L$,
the average chain length is given by $\bar{L}=1/p-1$ and the averaged
susceptibility by \cite{EggertAffleckHorton}
\begin{equation}
\label{Susci}
\chi_p =p^2\sum_L L(1-p)^L \chi(L) \; .
\end{equation}

\begin{figure}
\includegraphics*[width=0.99\columnwidth]{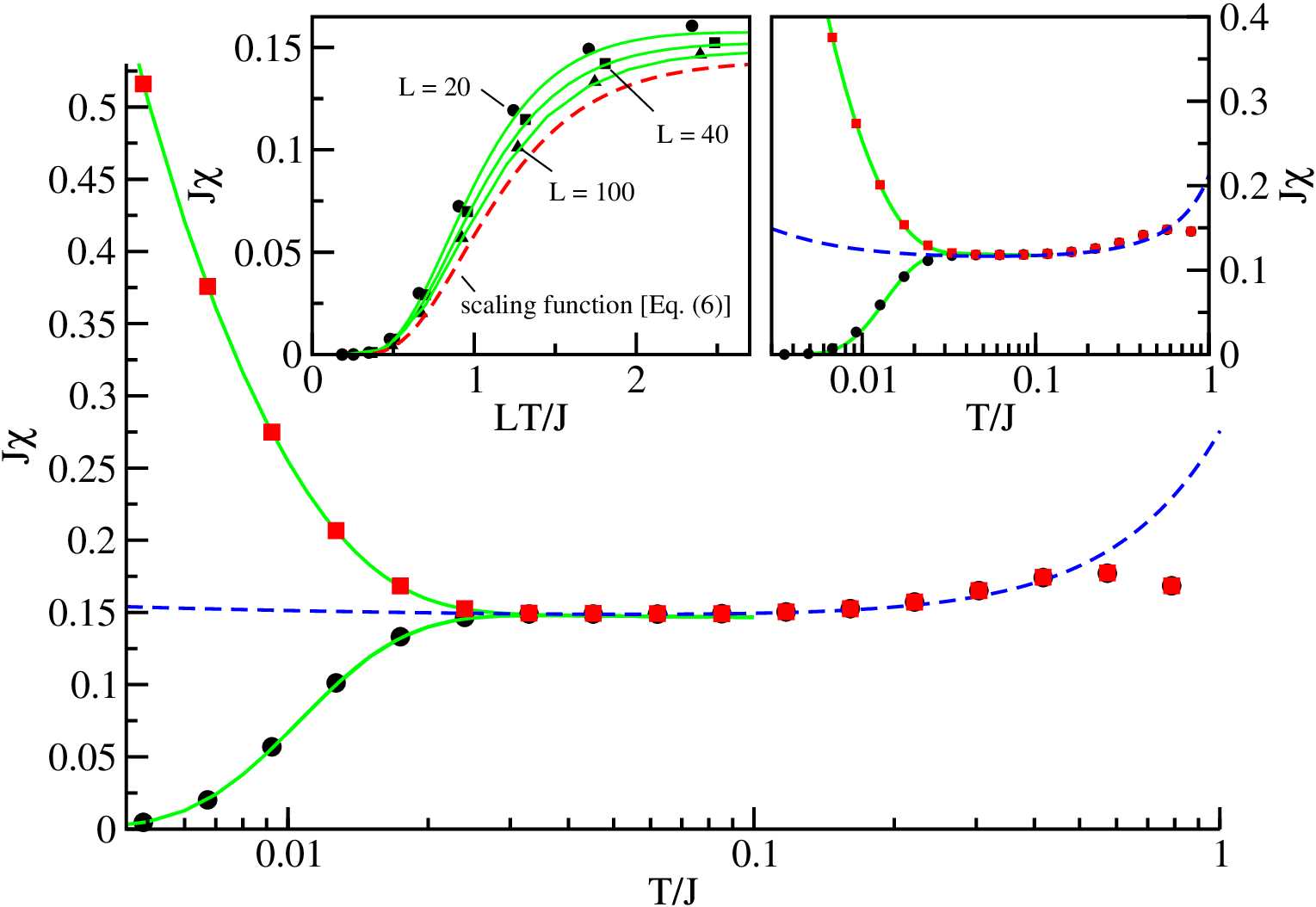}
\caption{(Color online) %% Comparison between 
  QMC data for $\Delta=3/4$ and $L=99$ (red squares), $L=100$ (black dots)
  compared to the field theory result derived here (green solid lines). The
  blue dashed line denotes $\chi=\chi_{\mbox{\tiny bulk}}+\chi_B/L$ (see
  Eqs.~(148, 149) in \cite{SirkerBortzJSTAT}). Left inset: Corrections to
  scaling for even length chains and $\Delta=3/4$. QMC data are represented by
  symbols, the field theory by the green lines. Right inset: Same as
  main figure for $\Delta=1$ with $a=5.8$ and $\chi_{\mbox{\tiny bulk}}\, ,
  \chi_B$ as obtained in \cite{Lukyanov,FujimotoEggert}. }

%% QMC data for $\Delta=3/4$ and different even (black dots) and odd chain
%%   lengths (red squares) and the field theory result derived in this letter
%%   (green solid lines) where the known results for $\chi_{\mbox{\tiny bulk}}$
%%   and $\chi_B/L$ (see Eqs.~(148, 149) in \cite{SirkerBortzJSTAT}) have been
%%   subtracted in each case. Inset: QMC data for $\Delta=1$ and $L=99$ (red
%%   squares), $L=100$ (black dots) and the field theory result with $a=14$
%%   (green lines). The blue dashed line is the field theory result for
%%   $\chi_{\mbox{\tiny bulk}}+\chi_B/L$ as obtained in
%%   \cite{Lukyanov,FujimotoEggert}.}
\label{fig3}
\end{figure}

We now calculate the susceptibility $\chi(L)=\langle (\sum_i
S^z_i)^2\rangle/(LT)$ in the low $T$ large $L$ limit using field-theory
methods. We start with the Hamiltonian (\ref{Hseg}) in the scaling limit,
i.e., ignoring irrelevant operators. The Hamiltonian is then equivalent to a
free boson model \cite{Affleck_lesHouches}
\begin{equation}
\label{BHam}
H =\frac{v}{2} \int_0^L dx \left[\Pi^2+(\partial_x\phi)^2\r] \; . 
\end{equation}
Here $\phi$ is a bosonic field obeying the standard commutation rule
$[\phi(x),\Pi(x')]=\im\delta(x-x')$ with $\Pi=v^{-1}\partial_t\phi$. The
velocity $v$ is a known function of the anisotropy $\Delta$. Using the mode
expansion for OBCs\cite{EggertAffleck92}
\begin{eqnarray}
\label{ModeExp}
\phi(x,t) &=& \pi R +2\pi R S_z \frac{x}{L} \\
&+& \sum_{n=1}^\infty
\frac{\sin\l(\pi nx/L\r)}{\sqrt{\pi n}}\l(\e^{-i\pi n \frac{vt}{L}}a_n +
\e^{i\pi n\frac{vt}{L}}a_n^\dagger\r) \nonumber
%% \e^{i\pi n\frac{vt}{L}}a_n^\dagger   
\end{eqnarray}
where $R$ is the compactification radius of the bosonic field, the Hamiltonian
(\ref{BHam}) can also be expressed as
\begin{equation}
\label{MHam}
H=\frac{\pi v}{KL}S_z^2-hS_z+\frac{\pi v}{L}\sum_{n=1}^\infty n(a_n^\dagger
a_n +1/2) \; .
\end{equation}
Here $a_n$ is a bosonic annihilation operator, $S_z$ an integer (half-integer)
for $L$ even (odd) and $h$ the magnetic field. $K$ - the so called Luttinger
parameter - is related to the compactification radius by $K=1/(2\pi R^2)$ and
is a known function of anisotropy $\Delta$ as well.  The susceptibility in the
scaling limit is then given by
\begin{equation}
\label{scalingpart}
\chi_s = -\frac{\partial^2 f}{\partial h^2}\bigg|_{h=0} 
=\frac{1}{LT} \frac{\sum_{S_z} S_z^2 \exp\l[-\frac{\pi
    v}{KLT}S_z^2\r]}{\sum_{S_z} \exp\l[-\frac{\pi v}{KLT}S_z^2\r]} \, .
\end{equation}
For $LT/v\to 0$ and $L$ even %% the susceptibility vanishes exponentially 
$\chi_s \sim \frac{2}{LT}\exp\l[-\frac{\pi v}{KLT}\r]$ whereas for $L$ odd
$\chi_s \sim (4LT)^{-1}$. For $LT/v\to \infty$ the thermodynamic limit result
$\chi_s = K/(2\pi v)$ is recovered. Corrections to scaling occur due to
irrelevant bulk and boundary operators. The leading bulk irrelevant
operator for %% in the antiferromagnetic regime 
$0<\Delta\leq 1$ is due to Umklapp scattering yielding the following
correction to (\ref{BHam})
\begin{equation}
\label{Umklapp}
\delta H =\lambda_1 \int_0^L dx \cos\l(2\phi/R\r) \, ,
\end{equation}
with $\lambda_1$ being the Umklapp scattering amplitude. 
%% In the following we
%% will consider this correction in first order perturbation theory for $L$ {\it
%%   and} $T$ finite. First and second order perturbation theory in this operator
%% in the thermodynamic limit $L\to\infty$ has been performed before. In first
%% order the leading contribution to the boundary susceptibility ($1/L$-term) is
%% obtained \cite{FujimotoEggert,BortzSirker,SirkerBortzJSTAT} whereas the second
%% order gives the leading temperature dependent correction to the constant bulk
%% susceptibility obtained from (\ref{scalingpart}) in the limit $L\to\infty$
%% \cite{Lukyanov}. 
To obtain $\chi$ to first order in the operator (\ref{Umklapp}) for $L$ {\it
  and} $T$ finite the expectation value $\langle\exp(\pm 2\im\phi/R)\rangle$
has to be calculated. Using the mode expansion (\ref{ModeExp}) this
expectation value splits into an $S_z$ (zero mode) and an oscillator part.
Upon using the cumulant theorem for bosonic modes we obtain $\langle\exp(\pm
2\im\phi/R)\rangle = \langle\exp(\pm
2\im\phi/R)\rangle_{S_z}\exp(-2\langle\phi\phi\rangle_{osc.}/R^2)$ with
\begin{equation}
\label{osc1}
\langle\phi\phi\rangle_{\mbox{\tiny osc.}} = \sum_{l=1}^\infty
\frac{\sin^2(\pi lx/L)}{\pi l}\l(1+\frac{2}{\e^{\pi vl/(TL)}-1}\r) \; .
\end{equation}
Introducing a cutoff $\alpha$ for the zero temperature part in (\ref{osc1})
and using $\sum_{l=1}^{\infty} z^l/l = -\ln(1-z)$ for $|z|<1$ we obtain the
following correction to %% the scaling form of the susceptibility in
(\ref{scalingpart}) 
\begin{eqnarray}
\label{suscicorr}
\delta\chi_1 &=&
\frac{2\tilde{\lambda}_1}{T^2}\l(\frac{\pi}{L}\r)^{2K}\!\!\eta^{6K}\l(\e^{-\frac{\pi
    v}{TL}}\r)  \int_0^{1/2} \!\!\!\!\!\! dy \,\frac{g_0\l(y,\e^{-\frac{\pi
      v}{KLT}}\r)}{\theta_1^{2K}\l(\pi y,\e^{-\frac{\pi v}{2TL}}\r)} \nonumber
\\
&\times& 2^{2K}\sin^{2K}\l(\pi y\r)\l[\l(1-\e^{2\pi\im
 y}\e^{-\alpha\pi/L}\r)\times \mbox{h.c.}\r]^{-K}  %%\l(1-\e^{-2\pi\im y}\e^{-\alpha\pi/L}\r)\r]^{-K}
\end{eqnarray}
with
\begin{eqnarray}
\label{suscicorr2}
g_0\l(y,q\r) &=& -\frac{\sum_{S_z} S_z^2\cos(4\pi S_z y) q^{S_z^2}}{\sum_{S_z}
  q^{S_z^2}} \\
&+&\frac{\l(\sum_{S_z} \cos(4\pi S_z y) q^{S_z^2}\r)\l(\sum_{S_z}
  S_z^2  q^{S_z^2}\r)}{\l(\sum_{S_z}
  q^{S_z^2}\r)^2} \, . \nonumber
\end{eqnarray}
Here $\eta(x)$ is the Dedekind eta-function, $\theta_1(u,q)$ the elliptic
theta function of the first kind \cite{Gradshteyn} and $\tilde{\lambda}_1
=\alpha^{2K}\lambda_1$. For $K<3/2$ the integral in (\ref{suscicorr}) is
convergent, the cutoff can be dropped, and the last line becomes equal to one.
%% In this case (\ref{suscicorr}) constitutes a universal
%% correction to the scaling form (\ref{scalingpart}).
%% we can drop the cutoff in the last line
%% leading to the following universal correction to scaling
%% \begin{eqnarray}
%% \label{suscicorr3}
%% \delta\chi_1 &=&
%% \frac{2\tilde{\lambda}_1}{T^2}\l(\frac{\pi}{L}\r)^{2K}\eta^{6K}\l(\e^{-\pi
%%     v/(TL)}\r) \\
%% &\!\!\!\!\!\!\!\!\times &\!\!\!\!\!\!\int_0^{1/2}dy \; g_0\l(y,\e^{-\pi v/(KLT)}\r)\theta_1^{-2K}\l(\pi
%%   y,\e^{-\pi v/(2TL)}\r) \; , \nonumber
%% \end{eqnarray}
%% where $\tilde{\lambda}_1 =\alpha^{2K}\lambda_1$. 
%% %% has been determined in \cite{Lukyanov} by comparing with the BA
%% solution. 
An additional correction arises due to the presence of an irrelevant boundary
operator. In essence, this operator changes the length of the chain $L$ to
some effective length $L+a$ \cite{EggertAffleck92}.  Replacing $L$ by $L+a$ in
the exponentials in (\ref{scalingpart}) and expanding to lowest order in $a$
yields
\begin{equation}
\label{NonU}
\delta \chi_2 = -\pi va\; g_1(\exp[-\pi
  v/(KLT)])/(KT^2L^3) 
\end{equation}
with 
\begin{equation}
\label{Taylor2}
g_1\l(q\r) = \frac{(\sum_{S_z} S_z^2 q^{S_z^2})^2}{(\sum_{S_z}
  q^{S_z^2})^2} -\frac{\sum_{S_z} S_z^4 q^{S_z^2}}{\sum_{S_z}
  q^{S_z^2}} \, . 
\end{equation}
For $K>3/2$ we isolate the cutoff dependence by subtracting the Taylor expand
of the integrand in (\ref{suscicorr}) to first non-vanishing order in $y$, and
then take $\alpha\to 0$, yielding a convergent integral. The cutoff dependent
part has exactly the same form as (\ref{NonU}). We can therefore think of $a$
in (\ref{NonU}) as a parameter incorporating both contributions. In the
thermodynamic limit $g_1\l(\e^{-\pi v/(KLT)}\r) \to K^2T^2L^2/(2\pi^2v^2)$ and
$\delta \chi_2 \to Ka/(2\pi v L)$. In this limit we can compare the field
theory result with a recent calculation of the boundary susceptibility based
on the Bethe ansatz \cite{BortzSirker,SirkerBortzJSTAT} leading to
\begin{equation}
\label{NonU3}
a=2^{-1/2}\sin\l[\pi K/(4K-4)\r]/\cos\l[\pi/(4K-4)\r] \; .
%% a=\frac{1}{\sqrt{2}}\frac{\sin\frac{\pi K}{4K-4}}{\cos\frac{\pi}{4K-4}} \; .
\end{equation}
Because the amplitude $\tilde{\lambda}_1$ is known exactly as well
\cite{Lukyanov,SirkerBortzJSTAT}, the obtained result is {\it parameter-free}.
In Fig.~\ref{fig3} a comparison between this field theory result and QMC data
for $\Delta=3/4$ is shown and in the left inset the corrections to the scaling
limit (\ref{scalingpart}) due to the leading irrelevant bulk and boundary
operator are visualized.
More generally, we always find excellent agreement if $T/J\lesssim 0.1$ and
$L\gtrsim 10$. The range of validity can be extended in principle by including
terms in higher order perturbation theory in $\lambda_1$. However, if $T/v\gg
1/L$ it is justified to separate $\chi$ into a bulk, $\chi_{\mbox{\tiny
    bulk}}$, and a boundary contribution, $\chi_B$, and to perform the
thermodynamic limit for each of them separately. A combination of this result
\cite{SirkerBortzJSTAT} (blue dashed line in Fig.~\ref{fig3}) with the result
derived in this letter yields an excellent parameter-free description of the
susceptibility for anisotropic chains of finite length up to $T\sim J/2$.

For $K=1$ ($\Delta=1$) perturbation theory alone is not sufficient because
Umklapp scattering becomes marginal. In this case the coupling constant
$\tilde{\lambda}_1$ has to be replaced by a renormalization group (RG)
improved running coupling constant \cite{AffleckGepner,Lukyanov}.
$\tilde{\lambda}_1$ then becomes a function of the two length scales $L$ and
$v/T$.
%% There is no general RG solution in this case. 
For small enough energies the smaller scale will always dominate and the
running coupling constant becomes $\tilde{\lambda}_1(L,v/T) = g/4$ with
\cite{Lukyanov}
\begin{equation}
\label{coup_iso}
1/g + \ln(g)/2 = \ln\l(\sqrt{2/\pi} e^{1/4+\gamma}\mbox{min}[L,v/T]\r) \; .
%% \tilde{\lambda}_1(T,\delta E)
%% =\frac{1}{4}\ln^{-1}\l(\frac{\Lambda}{\mbox{max}(T,\delta E)}\r)
\end{equation} 
where $\gamma$ is the Euler constant. $\delta \chi_1$ in the isotropic case is
therefore given by (\ref{suscicorr}) with the last line dropped and
$\tilde{\lambda}_1$ replaced by $\tilde{\lambda}_1(L,v/T)$. Furthermore, $K$
in (\ref{scalingpart}) should be replaced by $1+g(L,T)/2$ so that $KL$ in the
exponentials gets replaced by $L(1+g/2+a/L)$. For this case $a$ is not known
and we will use it as a fitting parameter.
%% For $\delta\chi_2$
%% (\ref{NonU}), $a$ in Eq.~(\ref{NonU3}) is not defined in the limit $K\to 1$
%% and we will use it as a length dependent fitting parameter. %% Note, that this parameter now
%% depends on $L$ and has to be determined for each length by comparing with
%% numerical data.
%% A comparison between QMC data and the field theory result in the isotropic
%% case is shown in the left inset of Fig.~\ref{fig3}. 
This way we obtain excellent agreement if $T/J\lesssim 0.1$ and $L\gtrsim 10$
(see right inset of Fig.~\ref{fig3}). At $LT\ll v$ the corrections to the
scaling formula (\ref{scalingpart}) in the isotropic case due to the marginal
interaction can also be understood as follows. In this limit the most
important correction to (\ref{scalingpart})
%% the most important
%% correction to the scaling formula (\ref{scalingpart}), due to the marginal
%% interaction, 
arises from the correction to the excitation energy of the lowest
excited states with $S^z=\pm 1$, $E\to (\pi v/L)[1-g(L)]$, \cite{QinAffleck}
corresponding to the replacement $1/K\to 1-g(L)$ in (\ref{scalingpart}). 
%% We
%% find that this simple replacement agrees surprisingly well with our QMC data
%% for $\chi$ for a large range of $L$ and $T$.  
Here $g(L)$ is given by (\ref{coup_iso}) but with min$[L,v/T]$ simply replaced
by $L$.
%% Semi-empirically we find that the replacement $K\to 1+g+\cdots$ does describe
%% the renormalization of the low-energy excitations well \cite{Schneider}. Using
%% Eq.~(\ref{scalingpart}) with $K \to 1+g$ yields a parameter-free analytic
%% formula for $\chi(L)$ even in the isotropic case which agrees surprisingly
%% well with our QMC data at low temperatures (see right inset in
%% Fig.~\ref{fig3}).

These analytical results, supplemented by numerical data for $L\lesssim 10$,
can be used to calculate the averaged susceptibility (\ref{Susci}) at low
temperatures.
%% To obtain the averaged susceptibility for
%% $\Delta=1$, which is most relevant for experiment, over a larger temperature
%% range we have decided, however, to use QMC data directly. 
In addition, we have calculated $\chi(L)$ numerically for all $L\in [1,100]$
and certain additional $L$ up to $L=1000$ at various temperatures. An
interpolation and extrapolation was then used to obtain $\chi(L)$ for all
lengths up to $L=5000$ at a given temperature \footnote{See EPAPS Document No.
  [] for the QMC as well as the interpolated and extrapolated data.}. This allows us
to reliably calculate $\chi_p$ (\ref{Susci}) for impurity
concentrations $p\gtrsim 0.2\%$ ($\bar{L}\lesssim 500$).
%% because the normalized
%% distribution of lengths $D_p(L)=p^2L(1-p)^{L-1}$ drops of rapidly for
%% $L\gg\bar{L}$ so that it is sufficient to consider only segments of length
%% $L\lesssim 10\bar{L}$.

Experimental data for the susceptibility of the quasi one-dimensional compounds
Sr$_2$Cu$_{1-x}$Pd$_x$O$_{3+\delta}$  have been analyzed 
in \cite{AmiCrawford,MotoyamaEisaki,KojimaYamanobe} by
decomposing it into $\chi=\chi_0 + \chi_C + \chi_{\mbox{\tiny inf.}}$.
 Here $\chi_0$ represents a
constant part due to core diamagnetism and Van Vleck paramagnetism, $\chi_C$
 a Curie term and $\chi_{\mbox{\tiny inf.}}$ the susceptibility 
(per unit length) of the
infinite $S=1/2$ Heisenberg chain. However, our results show 
that  non-magnetic impurities 
(chain breaks) have a quite different effect than paramagnetic impurities. 
\begin{figure}
\includegraphics*[width=0.79\columnwidth]{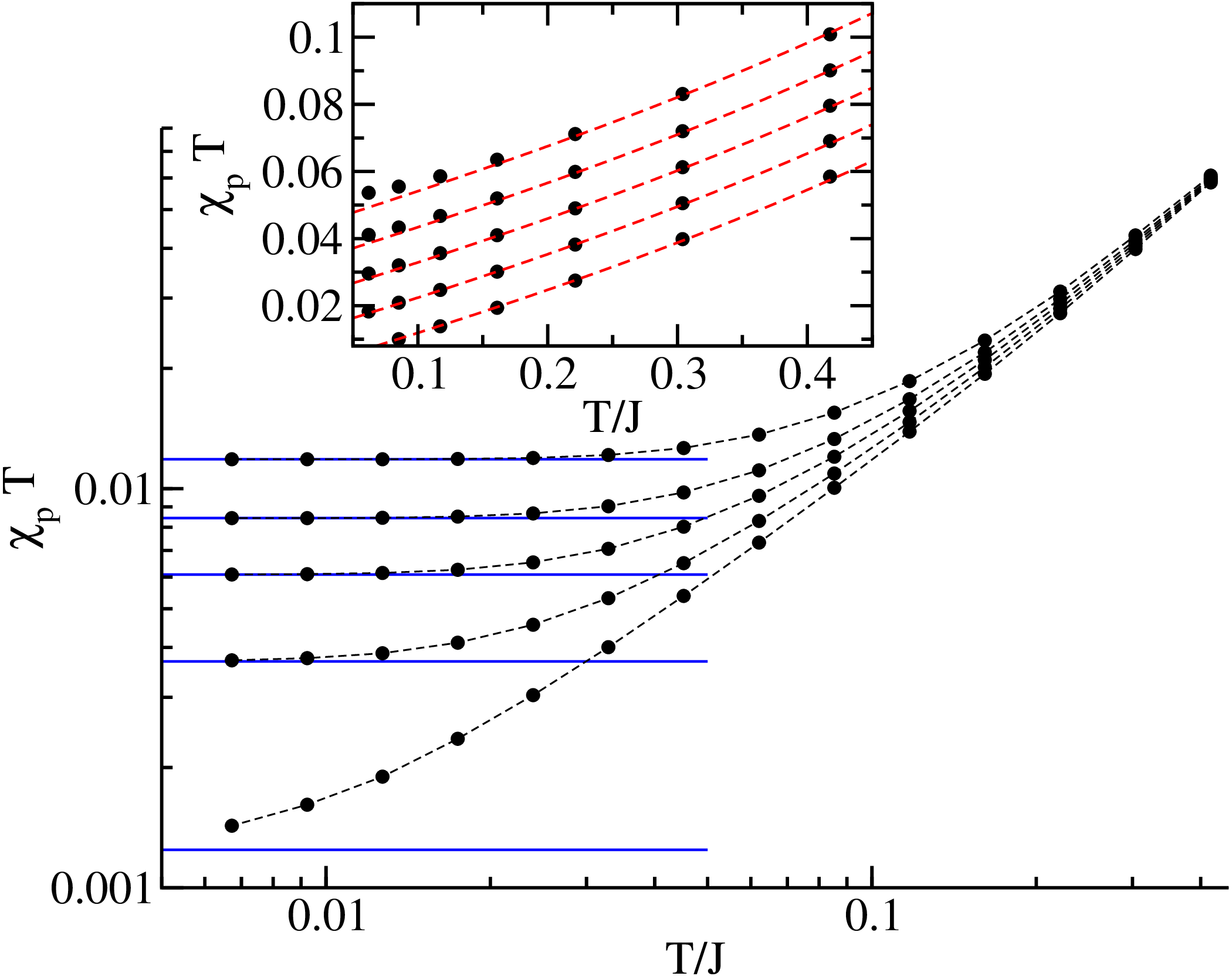}
\caption{(Color online) $T \chi_p$ for $p=0.01,0.03,0.05,0.07,0.1$ from bottom
  to top (black dots with dashed lines being a guide to the eye). The blue
  solid lines denote the value $(p/4)\cdot (1-p)/(2-p)$ reached in the limit $T\ll
  pv$.  Inset: Same data for $T \chi_p$ with subsequent curves shifted by
  $0.01$ compared to $T\chi_{\mbox{\tiny bulk}}+p/[12\ln(2.9J/T)]$ (red dashed
  lines) reached in the opposite limit.}
\label{fig1}
\end{figure}
%% In the extreme low $T$ limit, $T\ll pv$, the chain breaks will lead to
%% $1/T$-behaviour with a strength which is exactly half that which would arise
%% from the same density of paramagnetic impurities, since half of the chains
%% have odd length. 
In the extreme low-$T$ limit, $T\ll pv$, chain breaks will lead to
$1/T$-behavior with a strength reduced from that which would arise from the
same density of paramagnetic impurities by the fraction of chains of odd
length, $(1-p)/(2-p)\approx 1/2$. In the opposite limit $pv\ll T$ (but $T$
still $\ll J$), chain breaks lead to a correction to the pure result of
$\approx p/[12T\ln (2.9J/T)]$ corresponding to an ``effective paramagnetic
impurity concentration'' of $\approx p/[3\ln (2.9J/T)]$ which can be $\ll p$
for $T\ll J$.  This behavior is illustrated in Fig.~\ref{fig1}.

\begin{figure}
\includegraphics*[width=0.99\columnwidth]{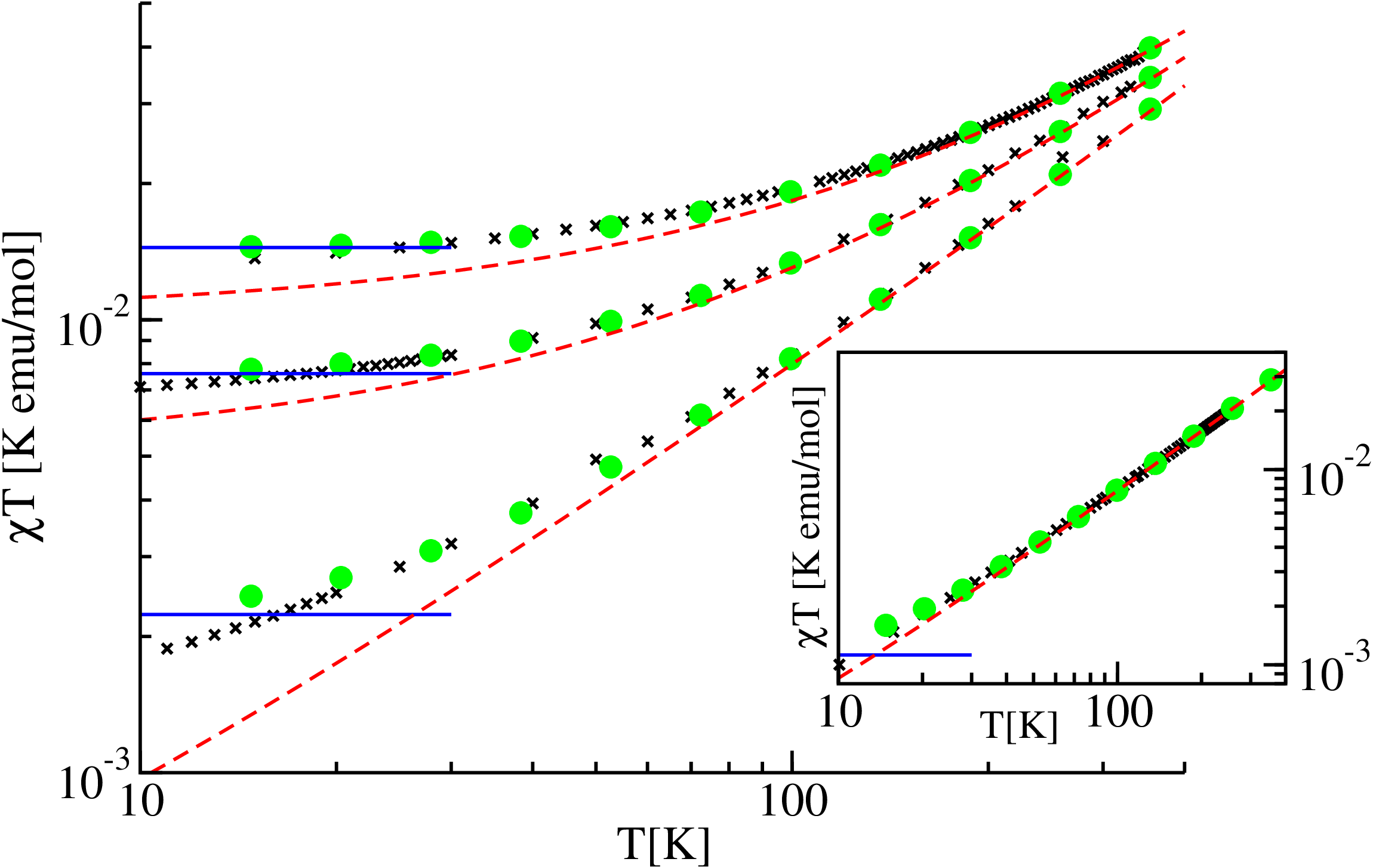}
\caption{(Color online) Measured susceptibility $T(\chi-\chi_0)$ for
  Sr$_2$Cu$_{1-x}$Pd$_x$O$_{3+\delta}$ with $x=0.5,\, 1,\, 3\%$ (crosses from
  bottom to top) where a constant $\chi_0$ (see Table \ref{tab1}) has been
  subtracted. The green dots represent the best theoretical fit with $p$ as in
  Table \ref{tab1}. Subsequent curves are shifted by $3\times 10^{-3}$.  The
  blue solid lines represent $(p/4)\cdot (1-p)/(2-p)$ and the red dashed lines
  $T\chi_{\mbox{\tiny bulk}}+p/[12\ln(2.9J/T)]$. Note that fits $\chi=\chi_0 +
  \chi_{\mbox{\tiny inf.}} + p_{\mbox{\tiny para}}/(4T)$ in the regime $T>50$
  K would be similar to the red dashed lines with effective paramagnetic
  impurity concentrations $p_{\mbox{\tiny para}}\approx p/[3\ln (2.9J/T)]$ an
  order of magnitude smaller than the nominal Pd concentrations.  Inset: Same
  for ``as grown'' sample of Sr$_2$CuO$_{3+\delta}$ from
  Ref.~\cite{MotoyamaEisaki}.}
\label{fig2}
\end{figure}

It has been shown that the Curie-like contribution to the susceptibility of
Sr$_2$CuO$_{3+\delta}$ can be significantly reduced by annealing under Ar
atmosphere indicating that this contribution is mostly due to excess oxygen
\footnote{From annealed samples we know that the true paramagnetic impurity
  concentration is $0.013\%$ at most \cite{MotoyamaEisaki} and
  can therefore be neglected in the following.}. We expect the additional
interstitial oxygen ions to be in a O$^{2-}$ ionization state. Each oxygen ion
would then dope two holes into the chains. If these holes are immobile they
effectively act like chain breaks. 
%% The interchain coupling has been estimated
%% to be only of order $J_\perp\sim 5$ K.  
It has been pointed out that the next-nearest neighbor coupling is not that
small, $J_2\sim J/16 \approx 140$ K \cite{RosnerEschrig}. Whereas we expect
this irrelevant coupling to cause only small corrections to the susceptibility
of an isolated chain segment the fact that it can bridge a chain break could
be significant.  A first order perturbative calculation, however, shows that
the correction to the averaged susceptibility is given by $\delta\chi_p\sim
pJ_2\chi_p^2$. We can therefore ignore both interactions as long as $T\gg
\mbox{max}(pJ_2,J_\perp)$ making the model discussed here applicable. As shown
in the inset of Fig.~\ref{fig2} it is indeed possible to explain the data in
\cite{MotoyamaEisaki} for the unannealed sample using a $0.6$\% concentration
of chain breaks.
If our picture is correct the amount of excess oxygen is significantly larger
than assumed in \cite{MotoyamaEisaki}, $0.06$\%.

The Sr$_2$Cu$_{1-x}$Pd$_x$O$_{3+\delta}$ samples used in \cite{KojimaYamanobe}
were not annealed. This complicates the analysis of the susceptibility data
because excess oxygen again leads to chain breaks in addition to those caused
by the non-magnetic Pd ions. The amount of excess oxygen hereby seems to vary
significantly from sample to sample. This is most apparent when comparing the
measured susceptibilities for $x=0.5\%$ and $x=1\%$ which are almost identical
apart from a small constant shift. We therefore treat the impurity
concentration $p$ as a fitting parameter. Results where we used $J=2200$ K and
a gyromagnetic factor $g=2$ are shown in Fig.~\ref{fig2} and Table \ref{tab1}.
\begin{table}[htbp]
%% \squeezetable
\caption{Concentration $x$ of Pd ions in experiment compared to impurity
  concentration $p$ and constant contribution $\chi_0$ yielding the best
  theoretical fit. The first line corresponds to the ``as grown'' sample of Sr$_2$CuO$_{3+\delta}$ from
  Ref.~\cite{MotoyamaEisaki}.}
%%  as well as the minimum triplet gap for the columnar dimer state in
%%  Fig.~\ref{fig4}(a)}
\label{tab1}
\begin{ruledtabular}
\begin{tabular}{ccc} %{rdddd}
\multicolumn{1}{c}{$x$ (Exp.)} &\multicolumn{1}{c}{$p$ (Theory)} 
&\multicolumn{1}{c}{$\chi_0$ [emu/mol]} \\
\hline\\[-0.3cm]
$0.0$ & $0.006$ & $-7.42\times 10^{-5}$ \\
$0.005$ & $0.012$ & $-7.7\times 10^{-5}$ \\
$0.01$ & $0.014$ & $-7.5\times 10^{-5}$ \\
$0.03$ & $0.024$ & $-7.5\times 10^{-5}$
\end{tabular}                                                       
\end{ruledtabular}                                                  
\end{table}
The figure clearly shows that the susceptibility data are indeed quite
different from what would be expected for paramagnetic impurities. Most of the
data are taken in the non-trivial crossover regime $pv\sim T$ and are well
described by our theory. Note, that the deviations at low temperatures are
expected because the suppression of the susceptibility due to interchain and
next-nearest neighbor interactions is not taken into account in our model. For
$x=0.005$ and $0.01$ the effective impurity concentrations $p$ are
larger than the nominal Pd concentration. $p-x$ roughly coincides with the
number of chain breaks in the ``as grown'' Sr$_2$CuO$_{3+\delta}$
%% (see inset of Fig.~\ref{fig2}) 
supporting our picture of additional chain breaks due to excess oxygen. 
%% Note
%% also, that $\chi_0$ in Table \ref{tab1} is almost independent from the Pd
%% concentration as expected corroborating the validity of our fits. 
For $x=0.03$, however, $p$ is smaller than the nominal Pd concentration.
Possible explanations are that for larger doping levels some of the Pd ions go
in interstitially instead of replacing Cu ions or that the sample exhibits
some sort of phase separation.
%% that their distribution is not completely random
%% \cite{SchmidtYushankhai}\footnote{$\chi_p$ can be calculated for any
%%  distribution function using our numerical data for $\chi(L,T)$ [18].}.
Future measurements on annealed Sr$_2$Cu$_{1-x}$Pd$_x$O$_3$ samples would be
helpful to clarify some of these issues.

\acknowledgments The authors acknowledge valuable discussions with
W.~N.~Hardy, K.~M.~Kojima and G.~A.~Sawatzky. This research was supported by
NSERC (J.S., N.L. I.A.), the DFG (J.S.) and the CIAR (I.A.). Numerical
simulations have been performed on the Westgrid network.

%% \bibliography{Literatur}

\end{document}